\documentclass[12pt,preprint]{aastex}
\usepackage{emulateapj5,apjfonts}
\usepackage{graphics}
\usepackage{amssymb}
\usepackage{bbm}
\usepackage{amsmath,textcomp,array}
\usepackage{onecolfloat}
\usepackage{bm}

\lefthead{Heitmann et al.}
\righthead{The Q Continuum Simulation}

\begin{document}


\def\head{

\title{The Q Continuum Simulation: Harnessing the Power of GPU
  Accelerated Supercomputers} 
\author{Katrin~Heitmann\altaffilmark{1,2}, Nicholas
  Frontiere\altaffilmark{1,3}, Chris Sewell\altaffilmark{4},
  Salman~Habib\altaffilmark{1,2}, Adrian Pope\altaffilmark{1,5}, 
Hal Finkel\altaffilmark{5}, Silvio Rizzi\altaffilmark{5}, Joe
Insley\altaffilmark{5}, Suman Bhattacharya\altaffilmark{1} } 

\affil{$^1$ HEP Division, Argonne National Laboratory, Lemont, IL 60439}
\affil{$^2$ MCS Division, Argonne National Laboratory, Lemont, IL 60439} 
\affil{$^3$ Department of Physics, University of Chicago, Chicago, IL 60637}
\affil{$^4$ CCS-7, CCS Division, Los Alamos National Laboratory, Los
  Alamos, NM 87545} 
\affil{$^5$ ALCF Division,  Argonne National Laboratory, Lemont, IL 60439}

\date{today}

\begin{abstract}

  Modeling large-scale sky survey observations is a key driver for the
  continuing development of high resolution, large-volume,
  cosmological simulations. We report the first results from the `Q
  Continuum' cosmological N-body simulation run carried out on the
  GPU-accelerated supercomputer Titan. The simulation encompasses a
  volume of $(1300~$Mpc$)^3$ and evolves more than half a trillion
  particles, leading to a particle mass resolution of $m_p\simeq
  1.5\cdot 10^8~$M$_\odot$. At this mass resolution, the Q Continuum
  run is currently the largest cosmology simulation available. It
  enables the construction of detailed synthetic sky catalogs,
  encompassing different modeling methodologies, including
  semi-analytic modeling and sub-halo abundance matching in a large,
  cosmological volume. Here we describe the simulation and outputs in
  detail and present first results for a range of cosmological
  statistics, such as mass power spectra, halo mass functions, and
  halo mass-concentration relations for different epochs. We also
  provide details on challenges connected to running a simulation on
  almost 90\% of Titan, one of the fastest supercomputers in the
  world, including our usage of Titan's GPU accelerators.

\end{abstract}

\keywords{methods: N-body ---
          cosmology: large-scale structure of the universe}}

\twocolumn[\head]

\section{Introduction}

The ever increasing depth, volume, and detail available in
observational catalogs from ongoing and future large-scale structure
surveys continues to emphasize the importance of large cosmological
simulations. This is particularly the case in the context of
`precision cosmology', where the simulations are a key source of the
essential theoretical predictions, especially in the nonlinear regime
of structure formation.

Cosmological simulations are used in many different ways: to
investigate new cosmological probes and their sensitivities, to test
analysis pipelines before applying them to real data, to understand
and model theoretical, astrophysical, and measurement systematics, to
obtain cosmological parameter constraints from the data, and to
estimate covariances -- the list is long.  In the absence of a
first-principles understanding of galaxy formation, and given the high
cost of carrying out simulations that include gas dynamics and
feedback effects, the only currently viable way to generate
sufficiently large synthetic sky catalogs is via gravity-only
simulations. The results from these simulations are then ``dressed
up'' with galaxies in post-processing, using a range of different
modeling approaches, such as halo occupation distribution (HOD)
(\citealt{kauffmann97}; \citealt{jing98}; \citealt{benson00};
\citealt{peacock00}; \citealt{seljak00}; \citealt{berlind02};
\citealt{zheng05}), Subhalo/Halo Abundance Matching (S/HAM)
(\citealt{vale04}; \citealt{conroy06}; \citealt{wetzel10};
\citealt{moster10}; \citealt{guo10}), or semi-analytic modeling (SAM)
(\citealt{white91}; \citealt{kauffmann93}; \citealt{cole94};
\citealt{somerville99}; \citealt{benson03}; \citealt{baugh06};
\citealt{benson10}). The details of these modeling approaches depend
strongly on the available mass and force resolution in the simulation
-- more resolution within halos and detailed tracking of halo
formation over time allow, in principle, for more sophisticated
modeling approaches, and therefore more detailed sky maps.

\begin{figure}[b]
\centerline{
 \includegraphics[width=3.5in]{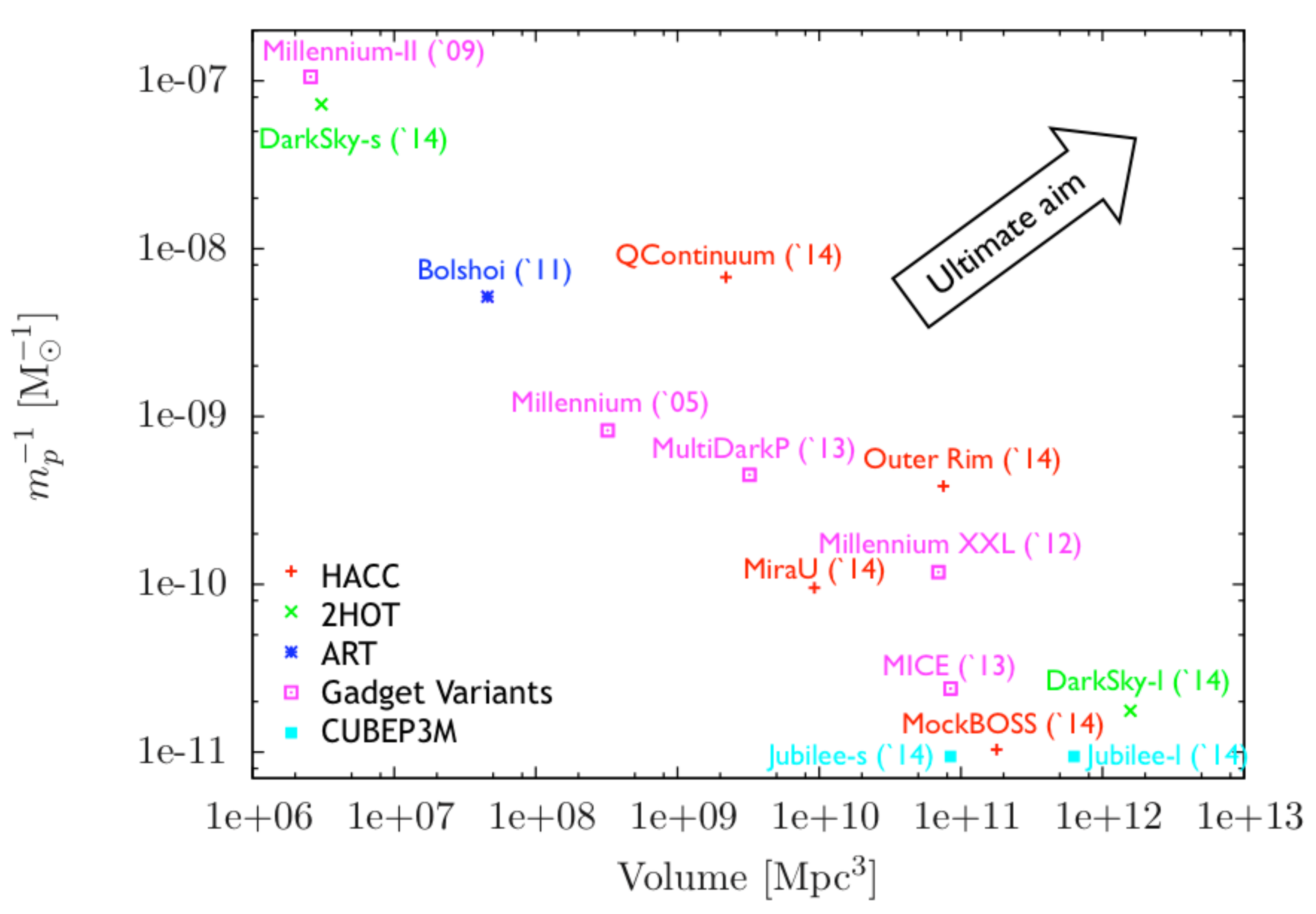}}
\caption{\label{sims} Representative selection of recent large
  cosmological N-body simulations with a minimum particle count of 8
  billion and mass resolution, $m_p\le 10^{11}$~M$_\odot$, where
  $m_p$ is the mass of a tracer particle in the simulation. }
\end{figure}

\begin{figure*}[t]
\centerline{
 \includegraphics[width=7.2in]{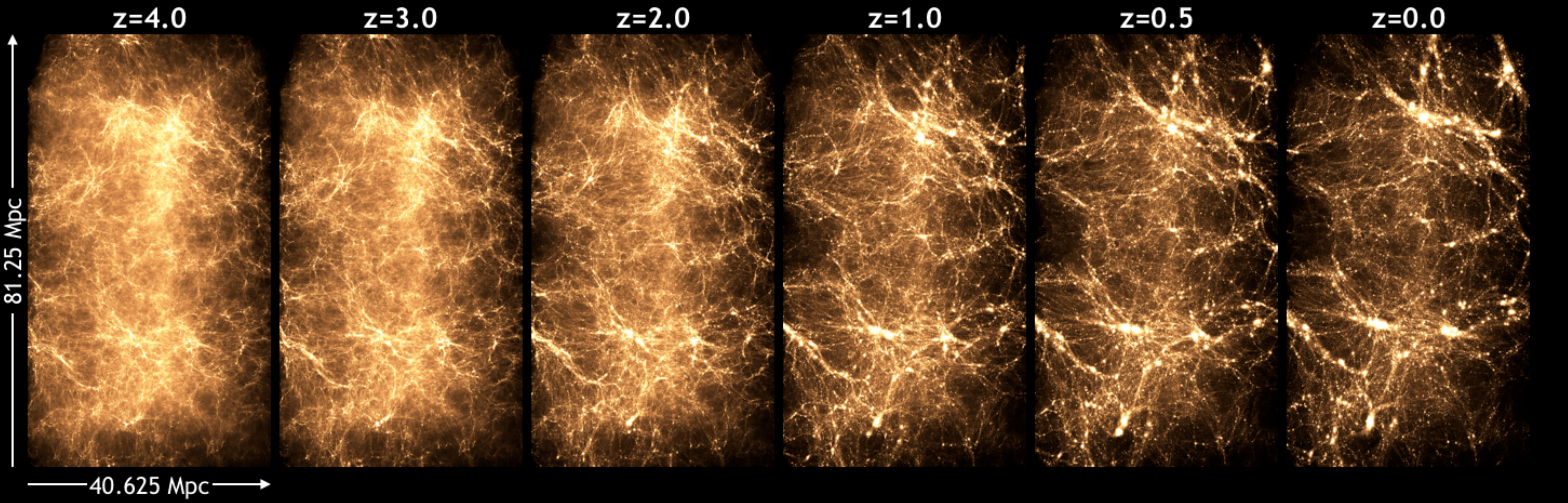}}
\caption{\label{evolve}Time evolution of the matter distribution
  between $z=4$ to $z=0$. Shown is the output from {\em one} of the
  16,384 nodes the simulation was run on. The visualizations here and
  in Figure~\ref{halop} are generated with the vl3 parallel volume
  rendering system~\citep{vl3} using a point sprite technique. }
\end{figure*}

Figure~\ref{sims} shows a selection of recently completed simulations
(this set is obviously incomplete and is presented only to convey a
qualitative impression of the current state of the art and where
cosmological simulation efforts are headed). We include
simulations that have at least a volume of $\sim (100~{\rm Mpc})^3$ (to
be relevant for cosmological studies), were run with at least $2048^3$
particles, and with particle masses not larger than
$m_p\sim10^{11}$~M$_\odot$ to enable the generation of reliable
synthetic sky catalogs. Besides the original
Millennium~\citep{springel05} and Millennium II~\citep{boylan09}
simulations, all runs have been carried out within the last three
years. Cosmological volume and the inverse particle mass (which
depends on the simulation volume, number of particles evolved, and
$\Omega_m$) are used as the master variables in the figure. The ideal
survey simulation aims to be as close to the upper right corner as
possible, characterized by a sufficiently large volume and with high
mass resolution. 

In Figure~\ref{sims} we have also attempted to capture the range of
N-body methods used in cosmology, including pure tree codes such as
2HOT used for the recent DarkSky simulations~\citep{skillman14},
hybrid tree particle-mesh (TPM) codes such as {\sc{Gadget}}-2 and
variants used for the Millennium simulations, the MultiDarkP
simulation\footnote{http://www.cosmosim.org/cms/simulations/multidark-project/mdpl/},
the MICE simulations~\citep{mice} as well as one of the HACC
implementations used for the Outer Rim simulation~\citep{habib14}, in
addition to particle-particle particle-mesh (P$^3$M) codes such as the
HACC implementation used in this paper for the Q Continuum simulation
and CUBEP3M used for the Jubilee suite~\citep{jubilee}, and finally
the adaptive mesh refinement-based code ART (Adaptive Refinement
Tree), used for the Bolshoi simulation~\citep{klypin}.

The computational costs for moving towards the upper right corner in
Figure~\ref{sims} are severe~\citep{power03}. Higher mass resolution
requires higher force resolution and increased time-resolution to
accurately resolve structures forming on small scales. For instance,
cluster-scale halos are resolved with many millions of particles (the
largest cluster in the Q Continuum simulation has more than 25 million
particles) adding significantly to the overall computational
costs. Another difficulty in increasing the size of simulations is the
available memory. Only the largest supercomputers possess sufficient
system memory to allow state of the art simulations. Unlike the
maximum available computational power, which continues to steadily
increase (although becoming harder to use), the available system
memory is not keeping pace with performance~\citep{kogge}.

To overcome current and future challenges posed by large cosmological
simulations with high mass and force resolution, we have designed the
HACC ({\bf H}ardware/Hybrid {\bf A}ccelerated {\bf C}osmology {\bf
  C}ode) framework, as described
in~\cite{habib09,pope10,habib14}. HACC runs very efficiently on all
currently available supercomputing architectures at full scale, and is
responsible for some of the largest cosmological simulations run to
date. In particular, HACC performs very well on accelerated systems,
such as Titan at the Oak Ridge Leadership Computing
Facility\footnote{https://www.olcf.ornl.gov/support/system-user-guides/titan-user-guide},
the machine used for the simulation described in this paper.  Titan is
comprised of 18,688 compute nodes, each containing a 16-core 2.2~GHz
AMD Opteron 6274 processor with 32~GB of RAM.  In addition, all of
Titan's compute nodes contain an NVIDIA Kepler accelerator (GPU) with
6~GB of local memory. This combination leads to more than 20~PFlops of
peak performance, enabling the Q Continuum simulation described
here. The Q Continuum simulation evolves more than 549 billion
particles in a (1300~Mpc)$^3$ volume. Figure~\ref{evolve} shows the
time evolution of the matter distribution from the output of a single
node (the full simulation run was carried out on 16,384 nodes),
covering a volume of $\sim$(81~Mpc $\times$ 81~Mpc $\times$
41~Mpc). These images give an impression of the detail in the matter
distribution as resolved by this simulation.

The Q Continuum run has been designed to address a number of
scientific targets. Because of its large dynamic range in both space
and mass, it allows accurate calculations of several quantities, such
as the mass power spectrum, the halo mass function, and the halo
concentration-mass relation, without having to resort to using nested
simulations. Another of its key scientific goals is the creation of
realistic synthetic sky maps for surveys such as
DES\footnote{http://www.darkenergysurvey.org/}~(Dark Energy Survey),
DESI\footnote{http://desi.lbl.gov/}~(Dark Energy Spectroscopic
Instrument), Large Synoptic Survey Telescope, LSST~\citep{lsst}, and
WFIRST-AFTA~\citep{wfirst}. In order to do this, halo/sub-halo merger
trees from finely-grained output snapshots will be used to generate
galaxy catalogs using the methods described above. The mass resolution
has been chosen to enable the construction of catalogs that can reach
useful magnitude limits as set by the survey requirements. Future
exascale supercomputers -- available on a timescale when LSST and
WFIRST will have turned on -- will provide sufficient computational
power to carry out simulations at similar or better mass resolution in
larger volumes, including subgrid modeling of a host of
`gastrophysical' effects.
 
The purpose of this paper is to describe the Q Continuum run,
including helpful details on how Titan's accelerators were used, and
to present a first set of validation and scientific results. Several
detailed analyses of the simulation outputs are currently underway and
will appear elsewhere. The paper is organized as follows. In
Section~\ref{hacc} we provide a very brief overview of HACC,
highlighting some of the GPU-specific implementation details used for
the Q Continuum run. The simulation itself and the outputs are
described in Section~\ref{QC_sim}.  We show results for selected
redshifts in Section~\ref{results} for the matter power spectrum, the
halo mass function, and the halo mass-concentration relation. We
conclude in Section~\ref{conclusion}, providing a summary of ongoing
projects carried out with the simulation.

\section{HACC and Accelerated Systems}
\label{hacc}

The development of the HACC framework was initiated in preparation for
Roadrunner's arrival at Los Alamos in 2008 -- the first machine to
break the Petaflop barrier. Roadrunner achieved its high performance
via accelerator units much in the same way as Titan achieves its high
performance today. The accelerators in Roadrunner's case were IBM
PowerXCell 8i Cell Broadband Engines (Cell BEs) compared to Titan's
GPUs. While the Cell BEs are very different from GPUs, the challenges
on the code design for these systems is basically the same: 1) memory
balance -- most of the memory on Titan resides on the (slow) CPU layer
of the machine, 2) communication balance -- while most of the
computation power resides in the accelerator units, the communication
bandwidth from the host to the accelerator can be a bottleneck, 3)
multiple programming models -- the accelerator units have very
specific language requirements, in the case of the GPU, either CUDA or
OpenCL (HACC uses the latter though a tree-PM GPU version written in
CUDA exists as well). A detailed description of how these challenges
are overcome in HACC is given in \cite{habib14}.

As is now standard for many cosmological N-body codes, the force
calculation in HACC is split into a long-range and a short-range force
solver. The long-range force solver is FFT-based and uses a spectrally
filtered particle-mesh (PM) method on a uniform grid. The PM solver is
the same on any architecture and dictates the weak scaling behavior of
HACC. Particular care has been taken to employ a very efficient
pencil-decomposed FFT implementation, specifically developed to work
well with the 3-d decomposition of the particle data
structures. HACC's short-range solver has been optimized for each
architecture encountered. On non-accelerated systems, HACC uses tree
algorithms, whereas on accelerated systems, adds solvers based on
direct $N^2$-methods that have been shown to also yield high
performance. The main trade-offs are complicated data structures as
they exist in tree codes versus compute intensive tasks in direct
particle-particle calculations. The range of implementations of HACC's
short-range solver allows for optimization across the different
architectures available and also allows for estimating the accuracy of
results by comparing simulations run with different force solvers.

The two key features that allow HACC to run efficiently across very
different supercomputing platforms (accelerated versus multi-core
systems such as IBM's BG/Q \citep{habib12}) are 1) an overload scheme
that enables straightforward on-node optimization of the short-range
solver, and 2) spectral filtering methods that push the hand-over
between the long-range and short-range solver to relatively small
length scales and therefore allow more compute-intense algorithms to
be implemented for the short-range solver. The original GPU code that
was run on Titan shortly after its arrival achieved more than
20~PFlops of peak performance (single precision) on $\sim 77\%$ of the
full machine and $\sim 8$~PFlops sustained performance earning HACC a
Gordon Bell finalist nomination in 2013~\citep{habib13}. The HACC
version used for the Q Continuum simulation run includes several
optimization schemes to further reduce the run time.

\subsection{HACC: Load-Balancing on the GPU}

The implementation of the HACC short-range force solver on GPUs has
been described elsewhere~\citep{habib13, habib14}. In addition, the
high-resolution simulation described in this paper relies on a
load-balancing scheme to take further advantage of the computational
power of the GPUs. While details about the implementation will be
provided elsewhere (Frontiere et al., in preparation) we briefly
explain the general idea here.

The principle behind the load balancing technique is to partition each
node volume into a set of overlapping data blocks, which contain
``active'' and ``passive'' particles -- analogous to the overloading
particle scheme between nodes.  Each block can independently perform
the short-range force calculations on its data, where it correctly
updates the interior active particles, and streams the bounding
passive particles. In this form, one can picture each node as a
separate HACC simulation, and the data blocks are the equivalent nodal
volume decompositions with overloading zones.  The scheme to perform a
short-range force timestep is as follows: 1) Each node partitions
itself into overlapping data blocks, 2) evolves the blocks
independently, and 3) reconciles the alive particles, whereby
removing the unnecessary duplicated passive ones.  Now that the
simulation data has been subdivided into smaller independent work
items, these blocks can be communicated to any nodes that have the
available resources to handle the extra workload. At this point, load
balancing can be performed, and one is limited only to the total
computation time of each block, as opposed to each node.  To perform
the actual load balancing, each node begins by globally sending each
processor the wall clock times of each of its owned blocks. The
timings are sorted to determine the fastest and slowest performing
nodes, and the theoretical balanced nodal time $T_{avg}$ is calculated
-- the total calculation time divided by the number of nodes. The
slowest node then calculates which block it would need to send to any
processor, such that the receiving processors total computation time
is as close to $T_{avg}$ as possible.  At this point the data is
communicated, and the receiving node's total time is incremented by
the block calculation time, whereas the sending node's total time is
subsequently decremented. This process is continued until no transfer
of data blocks will improve the total balanced time.  Once all of the
additional data blocks are received, each node performs the
short-range force calculation and returns the resulted data block to
its owner, in addition to communicating an updated timing for that
data block.

The data blocks form a 2-d decomposition, where the $x$-axis is
equally partitioned in PM grid units. As data is being duplicated to
partition the overlapping blocks, by definition, the total amount of
computation work has increased. This becomes worse and worse, the
finer the partitioning. However, the gains in load balancing can be
upwards of near a factor of ten, warranting the use of this scheme. We
have found that the optimal way to run is to partition the blocks to
be as big as possible on most nodes, and finely grain only the nodes
that are significantly taking more time than the average calculation
speed.  To properly balance the extra computation versus load balance
gain, the algorithm partitions data in blocks that are as large as
possible, unless its timing is more than a factor of two larger than
the average node speed. In this case, the data block is divided a
factor of two finer, and continues with such a partitioning until it
again becomes another factor of two out of balance. This allows for
adaptive load balancing, in addition to reducing the additional
computation as much as possible. The load-balancing scheme keeps the
computational time per step more or less constant even when entering
the highly clustered regime late into the simulation.

\subsection{Halo Analysis on the GPU}

\begin{figure}
\centerline{
 \includegraphics[width=3.5in]{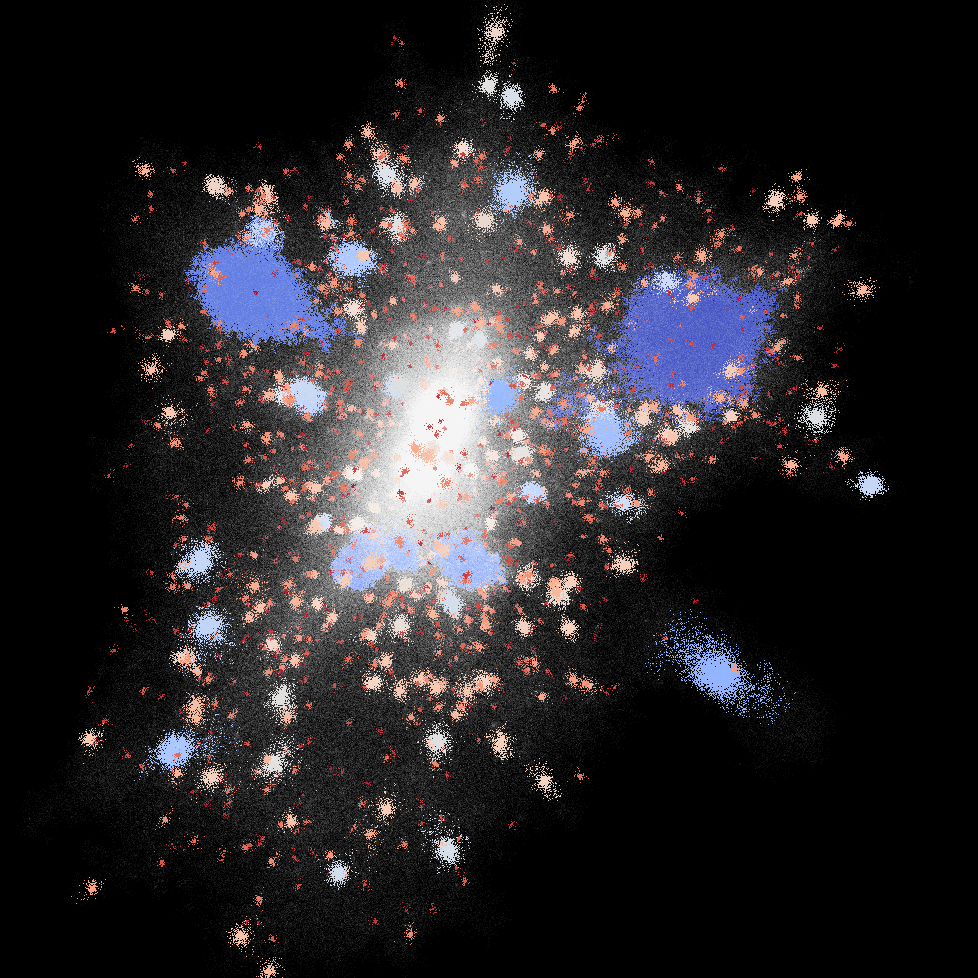}}
\caption{\label{albert}Substructure in a cluster-size halo of mass
  5.76$\cdot 10^{14}$M$_\odot$ at $z=0$. Subhalos are shown in
  different colors -- subhalo identification was run with a 20
  particle minimum per subhalo, resulting in 1364 subhalos. The light
  white background shows all the remaining particles that reside in the halo. }
\end{figure}

The computational effort expended on analysis of simulation runs as
large as the Q Continuum is a significant issue. One of the most time
consuming tasks is the generation of halo and subhalo catalogs. For
accelerated systems, the HACC philosophy has been to restrict the
analysis tasks to the CPUs because of portability concerns, given the
diverse nature of the typical analysis software suite. Usually the
analysis tasks take only a reasonably small fraction of the
computational time as compared to the actual N-body run, therefore the
extra effort for porting the analysis tools to the accelerators is not
justified. In the case of the Q Continuum simulation, however, the
analysis workload (due to the mass resolution and simulation size) is
large enough that the computational power of the GPUs needs to be
harnessed. Figure~ \ref{albert} shows the subhalo structure in a
cluster size halo with more than 4 million particles, as an example of
the complex structure of the halos that form in the simulation. (The
visualization was carried out using ParaView\footnote{ParaView --
  project webpage: http://www.paraview.org/}.)

The initial halo finding step (identifying particles that belong to a
friends-of-friends (FOF) halo) is still carried out on the CPU. As
detailed in \cite{woodring11}, the halo finder (based on a tree
algorithm) takes advantage of the overloaded data structure of the
main HACC code, as used for the long-range force solver. This allows
the halo finding to be carried out in an independently parallel
fashion, with a final reconciliation step to ensure that no halo is
counted more than once -- this requiring nearest neighbor
communication between MPI ranks. The CPU-based halo finding algorithm
is highly efficient. For the current simulation, finding all halos at
$z=2$ down to 40 particles per halo (resulting in 178,552,935 halos
holding $\sim 20\%$ of all the particles) took $\sim 6$ minutes, at
$z=1$ it took $\sim 10.4$ minutes to find 184,709,084 halos holding
$\sim 31\%$ of all the particles, and at $z=0$ it took just under 36
minutes to find 167,686,041 halos, holding $46\%$ of all the particles
in the simulation.

Following halo identification, a number of other tasks follow, such as
measurements of halo properties. One example of a computationally
expensive operation is the identification of the halo center as
defined by its potential minimum (``most bound particle''), especially
when dealing with halos that possess a large number of
particles. Accurate center-finding is important for measurements of
the halo concentration, for halo stacking (as for galaxy-galaxy
lensing), placing central galaxies from HOD modeling, etc. Although
our current potential estimation algorithm is well-suited for
accelerated hardware such as GPUs, we wished to avoid using custom
routines that would not be usable on alternative hardware. For this
reason we make use of the PISTON library~(\citealt{lo12}) to develop a
cross-architecture implementation, as described below.
 
It has been demonstrated that a wide variety of parallel algorithms
can be implemented efficiently using data-parallel
primitives~\citep{blelloch}.  Code written using this model can then
be easily ported to any hardware architecture for which
implementations of these data-parallel primitives exist.  The PISTON
library of visualization and analysis operators by \cite{lo12} and
\cite{sewell13} (part of the VTK-m project) is built on top of
NVIDIA's Thrust library, which provides CUDA, OpenMP, and Intel TBB
backends for data-parallel primitives\footnote{Thrust library --
  project webpage: http://thrust.github.io/}.  These primitives
include such operators as scan, transform, and reduce, each of which
can be customized with user-defined functors.  Thrust's API is very
similar to the C++ Standard Template Library (STL), in that it
provides vectors, iterators, and (parallel) algorithms that operate on
the vectors.  It also simplifies memory transfers by providing
host-side and device-side vectors, with easy copying between them.

By implementing a simple most bound particle center finder using
Thrust, we are able to make use of the GPUs on Titan without writing
code explicitly in CUDA.  Our code can also be compiled to Thrust's
OpenMP or Intel TBB backends to run on multi-core CPUs (including the
Xeon Phi).  While Thrust essentially functions as a source-to-source
translator, it may be possible to provide even more efficient support
for data-parallelism using, for example, compiler optimizations
\citep{jablin11}.  We use Thrust because it is a readily-available,
easy-to-use, production-quality, open-source, and effective library,
but our data-parallel algorithms should also be compatible with
alternative data-parallel frameworks. We tested the speed-up of the
center finder implemented on the GPU compared to the CPU on a
downscaled simulation (smaller volume but same mass resolution as the
Q Continuum run) at $z\sim 1.8$ and measured a speed-up of a factor of
$\sim 45$. More detailed timings on various platforms will be
presented elsewhere (Sewell et al., in preparation).
 
Since the distribution of particles is non-uniform, some nodes have
more halos and/or larger halos than others, particularly at later time
steps, thus making the work load unbalanced for center finding.
Rather than requiring many nodes with less work to wait idly while a
few nodes in very high-density regions complete their analysis, we
have utilized the streaming capability of HACC's customized GenericIO
library to quickly write particles of very large halos to disk, and
then stream these halos one at a time as a set of independent
single-node jobs that can be run on Titan or on another
GPU-accelerated machine, such as Moonlight at Los Alamos National
Laboratory. In future, a load balancing scheme will be implemented to
redistribute halos more evenly among the nodes for the center-finding
step during the initial {\em in situ} analysis run.

\subsection{I/O Performance on the Titan Lustre File System}

When attempting to carry out a simulation on large supercomputers, an
efficient I/O strategy is crucial. To put the size of the Q Continuum
run in context, the original Millennium simulation ~\citep{springel05}
stored 64 full time snapshots leading to 20TB of data. Here we store
101 time snapshots with more than 50 times as many particles per
snapshot, leading to a raw output of approximately 2~PB, a factor of
100 increase compared to the Millennium simulation. In addition,
check-points (full particle dumps including overloaded particles to
enable efficient restarts) need to be temporarily stored. The run time
of one simulation segment on Titan was optimally 24 hours (the maximum
length of the queue for production runs on the machine). To guard
against major computational time loss due to unexpected failures
during these 24 hour periods, we stored check-points every five to six
hours to enable intermediate restarts when necessary. Both raw outputs
and check-pointing need fast write and read routines.

For HACC we have developed a customized I/O strategy that allows us to
obtain almost peak I/O performance with relatively small adjustments
to the file system and the I/O node structure available. A detailed
description of our I/O implementation, called GenericIO, is given
in~\cite{habib14}.  Titan uses the Lustre file system, and while the
general I/O scheme and file format are the same across different file
systems, the way that the data from each MPI rank is divided into
individual output files is slightly different on different file
systems and optimized for each separately. Titan's Lustre
configuration uses approximately 1000 Lustre object storage targets
(OSTs) and each file is striped across 4 OSTs. To maximize the I/O
bandwidth, we distribute the data in an approximately-uniform manner,
such that the number of files multiplied by the stripe count
approximately equals the total number of OSTs. Thus, we write to a
total of 256 files to satisfy this condition. Ranks are divided evenly
among these files simply by dividing the MPI\_COMM\_WORLD rank number
by the total number of files (256) and using the remainder of that
operation as the rank's file number. This simple adjustment led to a
factor of 3 speed up in the I/O over the non-optimized implementation
(which was a random assignment of ranks to 256 output files),
to yield $\sim 36,000$~MB$/s$ for writing the raw output in less than
10 minutes and the checkpoint files in less then 20 minutes. The
reading of the files was faster by approximately $40\%$.

\section{The Q Continuum Simulation}
\label{QC_sim}

\begin{figure*}[t]
\centerline{
 \includegraphics[width=2.35in]{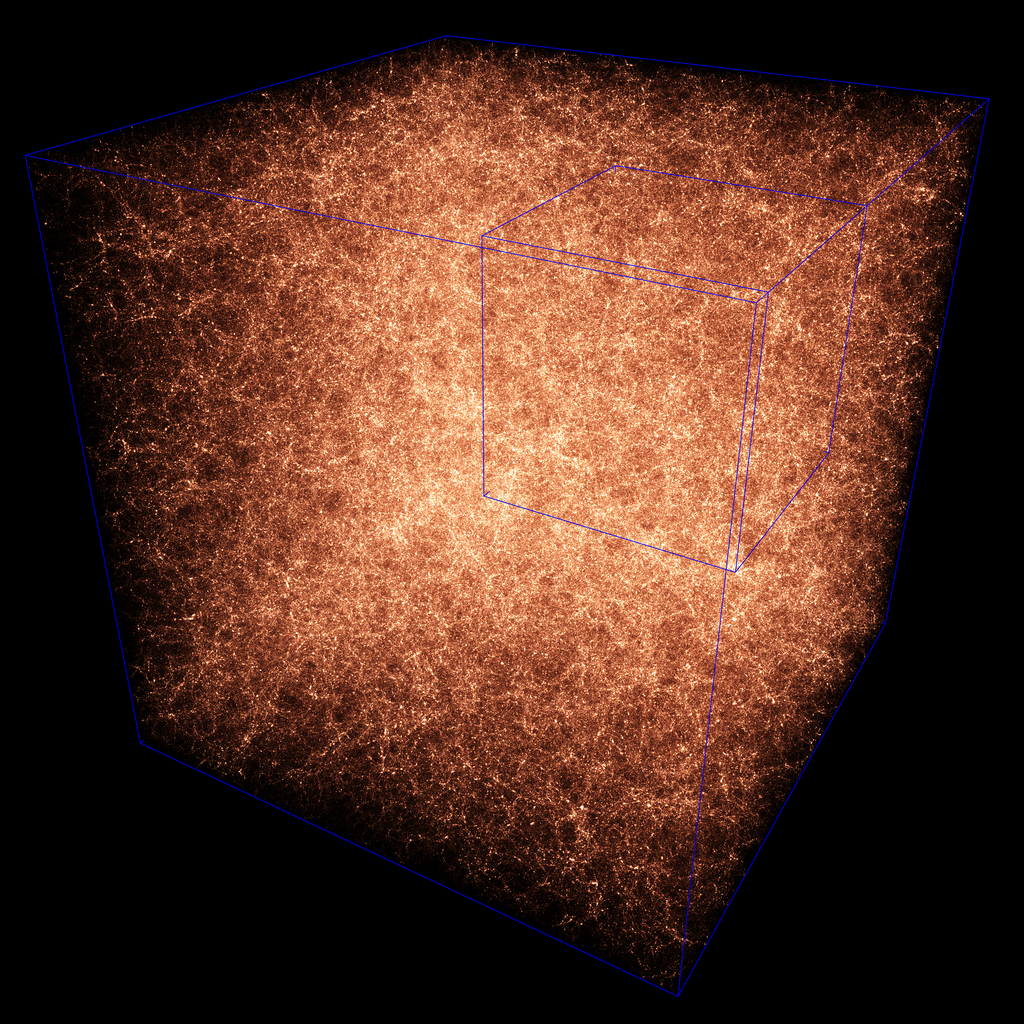}
 \includegraphics[width=2.35in]{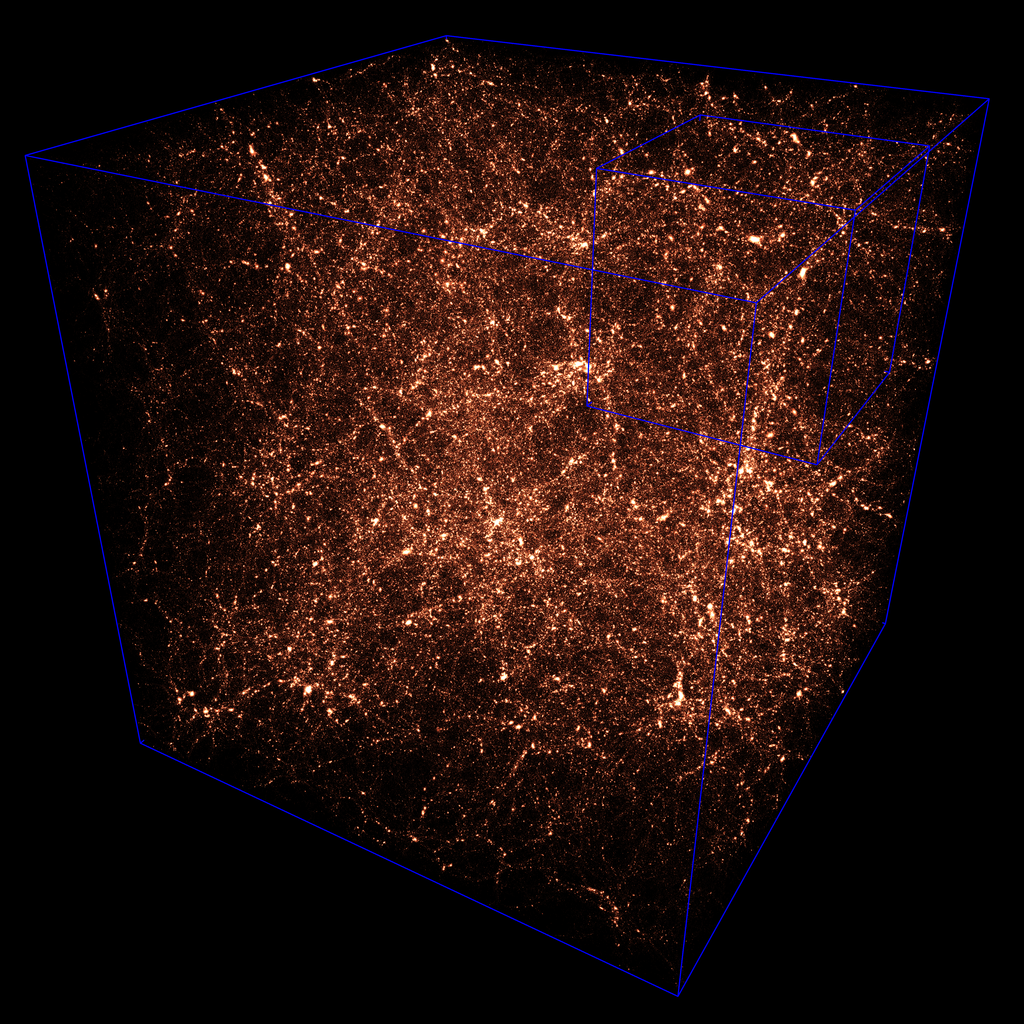}
 \includegraphics[width=2.35in]{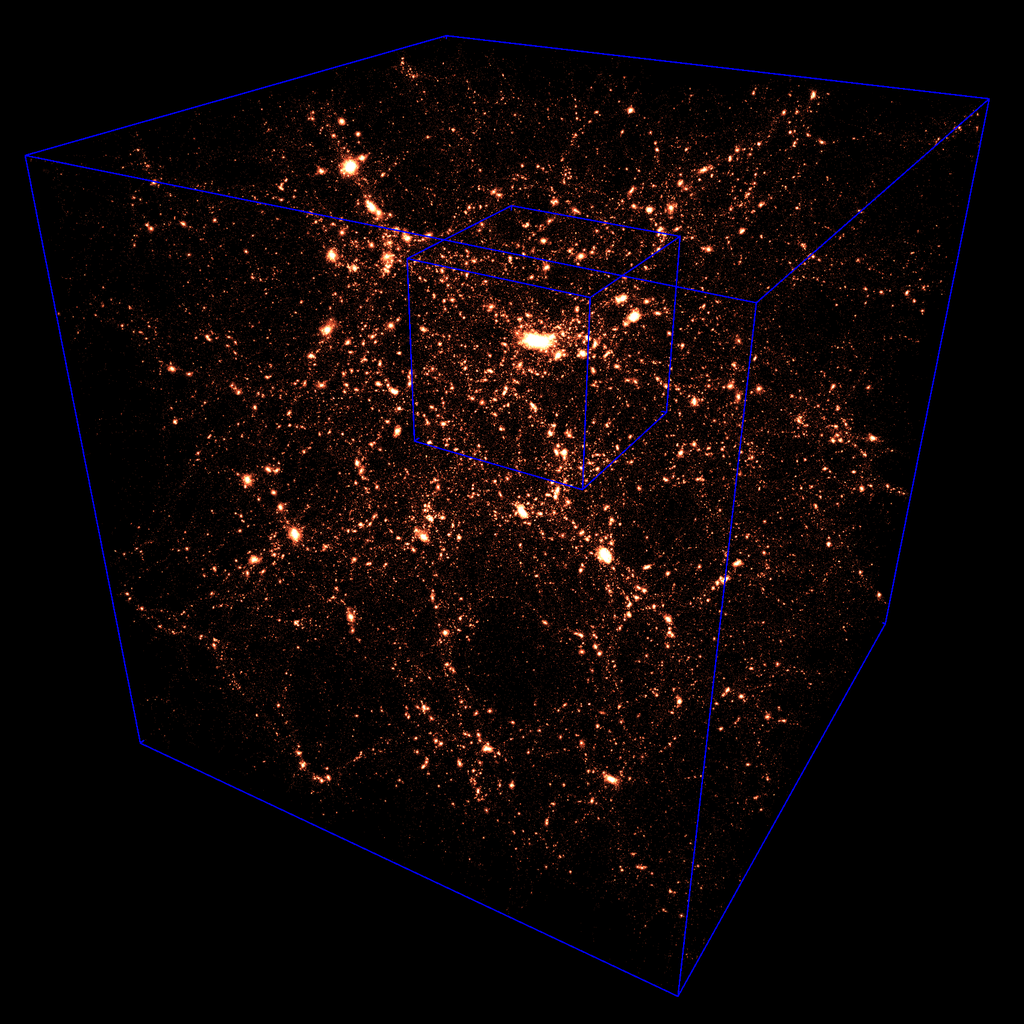}}
 \vspace{0.5mm}
 \centerline{
 \includegraphics[width=2.35in]{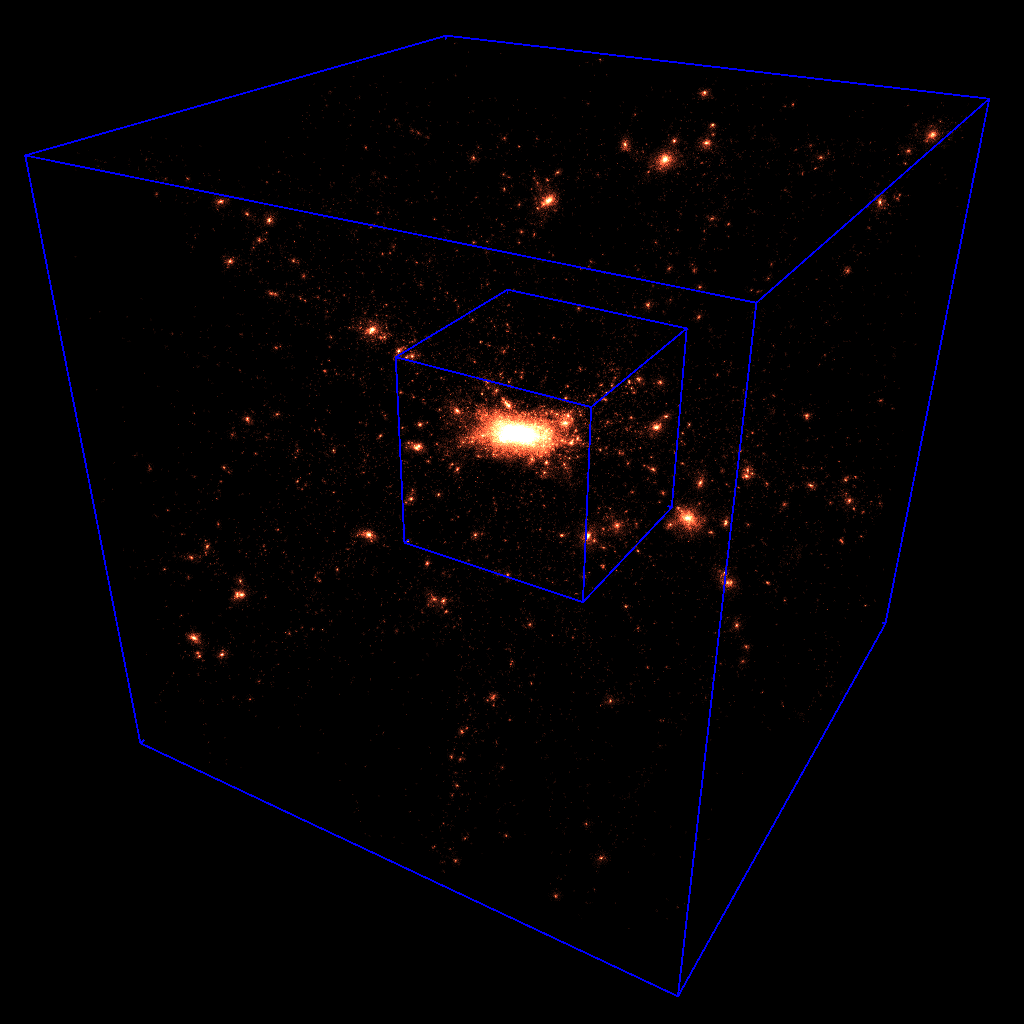}
 \includegraphics[width=2.35in]{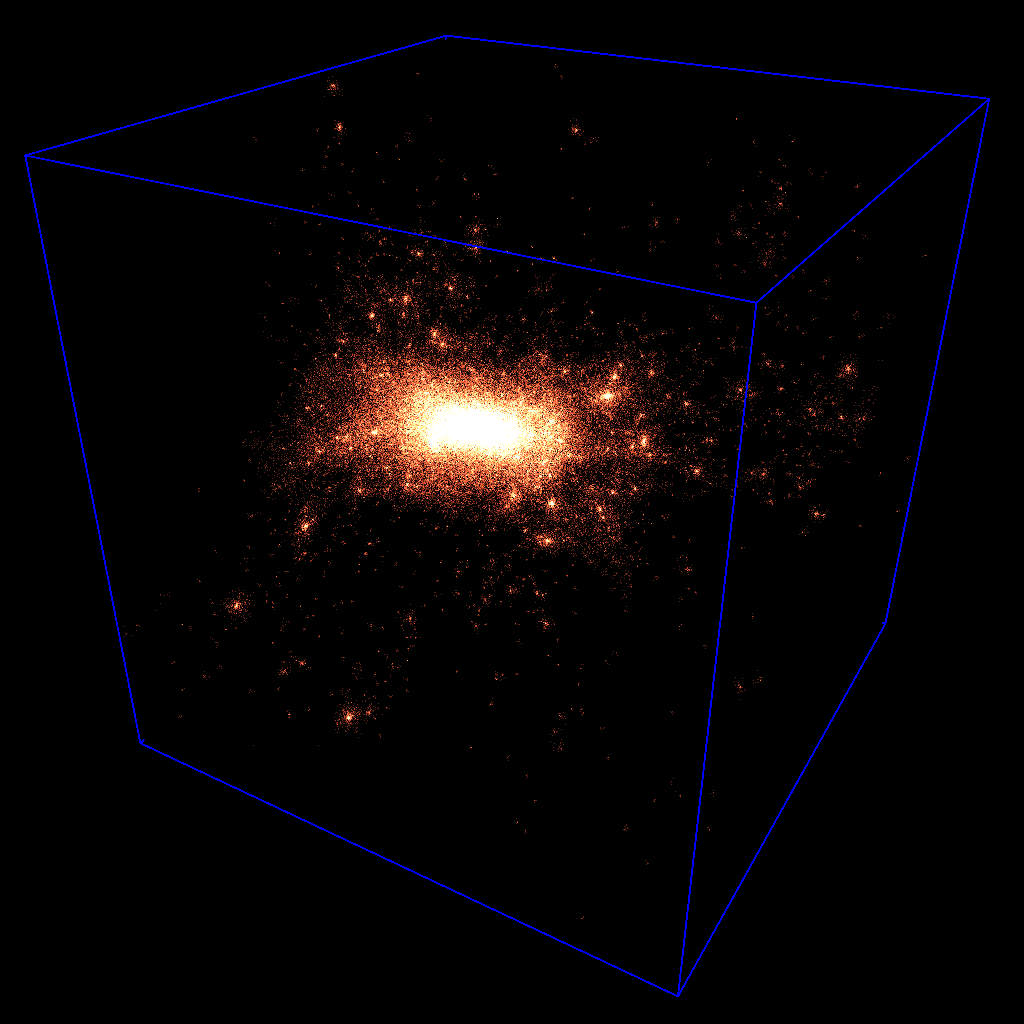}
 \includegraphics[width=2.35in]{bh4096.pdf}}
\caption{\label{halop}Halo particles in the Q Continuum simulation at
  $z=0$. Shown are a sub-sample of 1\% of all particles in halos of
  mass $1.48\cdot 10^{11}$M$_\odot$ and above. Halos with less than
  500 particles are represented by 5 particles. The first image (left
  upper corner) shows the full volume; the inset box volume is shown
  in the next image (upper middle) and so on. The series of images
  zooms into one of the very largest clusters in the simulation, shown
  in the last two images (lower panels, middle and right).  Overall,
  approximately 46\% of all particles reside in halos with 100 or more
  particles at $z=0$.}
\end{figure*}

In this section we provide details about the simulation parameters as
well as the outputs that have been stored. The simulation was carried
out on 16,384 nodes of Titan, utilizing almost 90\% of the full
machine. Details of the code implementations and optimizations and
code scaling results will be presented elsewhere (Frontiere et al., in
preparation).

Figure~\ref{halop} shows a visualization of the full simulation box as
well as zoom-ins to one of the most massive clusters in the simulation
at $z=0$. The images show $1\%$ of the particles that reside in halos
in order to highlight the cosmic web structure as seen in the
simulation. At $z=0$, there are more than 167 million halos with at
least 100 particles per halo. The cluster in the last image has a mass
of $\sim 3.75\cdot 10^{15}$~M$_\odot$, including more than 25 million
particles.

\subsection{Parameters}
The simulation is carried out using the best-fit cosmology as measured
by WMAP-7~\citep{wmap7}. The chosen cosmological parameters are:
\begin{eqnarray}
\omega_{\rm cdm}&=&0.1109\stackrel{h=0.71}{\Rightarrow}\Omega_{\rm
  cdm}=0.220\\ 
\omega_{\rm b}&=&0.02258,\\
n_s&=&0.963,\\
h&=&0.71,\\
\sigma_8&=&0.8,\\
w&=&-1.0\\
\Omega_{\nu}&=&0.0.
\end{eqnarray}
This cosmology has also been used for the Outer Rim simulation
~(\citealt{habib14}) and the Mira Universe~(\citealt{emu_ext}), resulting
in a set of simulations that cover different mass resolutions and volumes.
This provides a powerful testbed for answering questions related to convergence
of observable quantities with respect to resolution and volume.

The box size of the Q Continuum simulation is $L=1300~{\rm Mpc}$ $=$
$923~h^{-1}{\rm Mpc}$. The run employed $8192^3= 0.55$~trillion
particles, leading to a particle mass of
\begin{equation}
m_p=1.48\cdot 10^{8}~{\rm M}_\odot=1.05\cdot 10^8~h^{-1}{\rm M}_\odot.
\end{equation}
This is a good compromise for resolving halos and subhalos and still
covering a large cosmological volume. The force resolution was
2~$h^{-1}$kpc. The starting redshift for the simulation is
$z_{in}=200$ and the initial conditions are generated using the
Zel'dovich approximation~\citep{Zel70}. The transfer function is
obtained from {\tt CAMB}~\citep{camb}. We generated several initial
conditions, measured the initial power spectrum for each realization
and picked one that traces the baryonic wiggles on large scales well
(for the power spectrum results see Section~\ref{powspec}).

\subsection{Outputs}

We store 101 time snapshots between $z=10$ and $z=0$, evenly spaced in
$\log_{10}(a)$. This leads to the following output values in redshift:
\begin{eqnarray}
z&=&\left\{10.04, 9.81, 9.56, 9.36, 9.15, 8.76, 8.57, 8.39, 8.05,
  7.89, \right.\nonumber\\ 
&&7.74, 7.45, 7.31, 7.04, 6.91, 6.67, 6.56, 6.34, 6.13, 6.03, 5.84,
\nonumber\\ 
&&5.66, 5.48, 5.32, 5.24, 5.09, 4.95, 4.74, 4.61, 4.49, 4.37, 4.26,
\nonumber\\ 
&&4.10, 4.00, 3.86, 3.76, 3.63.  3.55, 3.43, 3.31, 3.21, 3.10,
3.04,\nonumber\\ 
&& 2.94, 2.85, 2.74, 2.65, 2.58, 2.48, 2.41, 2.32, 2.25, 2.17, 2.09,
\nonumber\\ 
&&2.02, 1.95, 1.88, 1.80, 1.74, 1.68, 1.61, 1.54, 1.49, 1.43, 1.38,
\nonumber\\ 
&&1.32, 1.26, 1.21, 1.15, 1.11, 1.06, 1.01, 0.96, 0.91, 0.86, 0.82,
\nonumber\\ 
&&0.78, 0.74, 0.69, 0.66, 0.62, 0.58, 0.54, 0.50, 0.47, 0.43, 0.40,
\nonumber\\ 
&&0.36, 0.33, 0.30, 0.27, 0.24, 0.21, 0.18, 0.15, 0.13, 0.10, 0.07,
\nonumber\\ 
&&\left.0.05,0.02, 0.00\right\}.
\end{eqnarray}
In addition to the raw time-snapshot data we also store several
analysis files as detailed below.

\subsubsection{Particle Outputs}

A full particle output requires roughly 20~TB of storage. Currently,
we have stored the full raw particle data from all 101 snapshots. In
the longer term, this data will be kept only on tape -- the storage
requirements of $\sim$2~PB make it prohibitive to store the data on
disk for an extended period of time. In addition to the full
snapshots, we have stored 1\% of all particles for 101 time snapshots
-- this will enable the calculation of the correlation functions and
high-resolution power spectra for the matter distribution later
on. Generating this down-sampled output on the fly saves considerable
compute resources since reading the full particle outputs back in
requires a large fraction of the supercomputer.

\subsection{Halo Outputs}

In this paper, we focus our analysis on a handful of halo snapshots at
$z\simeq 0,1,2$, found with an FOF halo finder with a linking length
of $b=0.2$. This choice leads to halo mass results closest to a
``universal" mass function and has been extensively studied (see,
e.g., \citealt{J01,reed06,cohn07,lukic07,tinker08,crocce09,courtin10,
  bhattacharya11}). It therefore offers a well-studied test case for
simulation results, along with quantities such as the mass power
spectrum. For the full set of outputs, we also store halo information
for a linking length of $b=0.168$ since this choice suffers less from
overlinking problems and is commonly used to create HOD based mock
catalogs (see, e.g., \citealt{reid11} for redshift space distortion
studies or \citealt{white10} for measurements of the clustering of
massive galaxies in the BOSS survey). Below we provide a list of the
outputs we store from this first level of analysis.

\begin{itemize}
\item FOF halo catalogs -- there are many scientific applications for
  FOF catalogs, such as building HOD-based galaxy catalogs, measuring
  mass functions, etc. For each halo we store: the number of particles
  in the halo, the halo tag, the halo mass, the halo center and
  velocity (mean and potential minimum), and the velocity
  dispersion. We store FOF halo information for halos with at least 40
  particles. Our fiducial linking length is $b=0.168$, for a small
  subset of the snapshots we also store halo files obtained with a
  linking length of $b=0.2$ to investigate the universality of the
  mass function as a function of redshift.

\item Particle and halo tags for all particles with long integers for
  all time snapshots -- this is to enable building detailed merger
  trees. These merger trees will be used to generate synthetic sky
  maps in post-processing using Galacticus, a semi-analytic model
  developed by~\cite{benson}.

\item All particle information for halos with more than 10,000
  particles, leading to a halo mass of $\simeq 1.5\cdot
  10^{12}~h^{-1}$M$_\odot$ and larger. These halos will be useful for
  studying the evolution of substructure, shapes of group and cluster
  sized objects, etc.

\item 1\% of particles in each halo, with a minimum of 5 particles per halo 
-- this output will be used to place galaxies into halos
  following HOD prescriptions.

\item Spherical overdensity (SO) halo catalogs -- again, these have
  many scientific applications. We will store halo catalogs for halos
  with 1000 particles and more. Currently, for each halo we store 21
  entries (FOF mass, tag, center position and velocity, SO mass and
  particle count and three different definitions of center, and SO
  radius). Storing both FOF and SO information in one file enables
  very easy evaluation of how relaxed the halos are by comparing the
  two mass definitions. Since we are only storing this information for
  large halos, the storage overhead is not very high.

\item SO profiles -- these outputs are used for deriving predictions
  for a number of quantities, such as the concentration-mass
  relation. For each halo we store information from 20 radial bins and
  the FOF tag, SO bin number, SO bin count, SO bin mass, SO bin
  radius, SO bin density, SO bin density ratio, and the SO bin radial
  velocity.

\end{itemize}

\section{Results}
\label{results}

We discuss results for three basic N-body simulation output
measurements, the matter power spectrum, the halo mass function, and
the halo concentration-mass relation. The large dynamic range of the Q
Continuum simulation allows us to validate previous results that came
from combining multiple simulations; these include verifying the
accuracy of determining the halo mass function and the construction of
emulation schemes, both derived from nested simulations. Given the
accuracy levels desired by near-future observations, these tests are
an essential component of any precision cosmology simulation campaign.

\subsection{Matter Power Spectrum}
\label{powspec}

In this section we show results for the matter power spectrum $P(k)$
at three different redshifts, $z=0,1$, and $z=2$. The power spectrum
is determined on the fly during the simulation and also serves as a
sensitive measurement to monitor the health of the run (for detailed
HACC power spectrum tests see~\cite{habib14}).  For a detailed
description of the power spectrum routine and accuracy controls
for power spectrum predictions, see, e.g., \cite{coyote1}. In summary,
we deposit the simulation particles onto a grid using a cloud-in-cell
(CIC) assignment. The application of a discrete Fourier transform then
yields $\delta(\bf k)$. From this we can determine $P(\bf
k)=|\delta(\bf k)|^2$ which is then binned in amplitudes to obtain
$P(k)$.  We compensate for the smoothing induced by the CIC assignment
by dividing $P(\bf k)$ by the window function that corresponds to the
CIC assignment scheme. When determined on the fly, the FFT size for
calculating the power spectrum is currently set to the PM force grid;
in the case of the Q Continuum simulation this is a grid of size
$8192^3$. Keeping the FFT grid for the power spectrum at this size
allows for fast evaluation of $P(k)$ -- only about a minute, roughly
equally spent between the CIC and the FFT evaluations.

\begin{figure}[t]
\centerline{
 \includegraphics[width=3.7in]{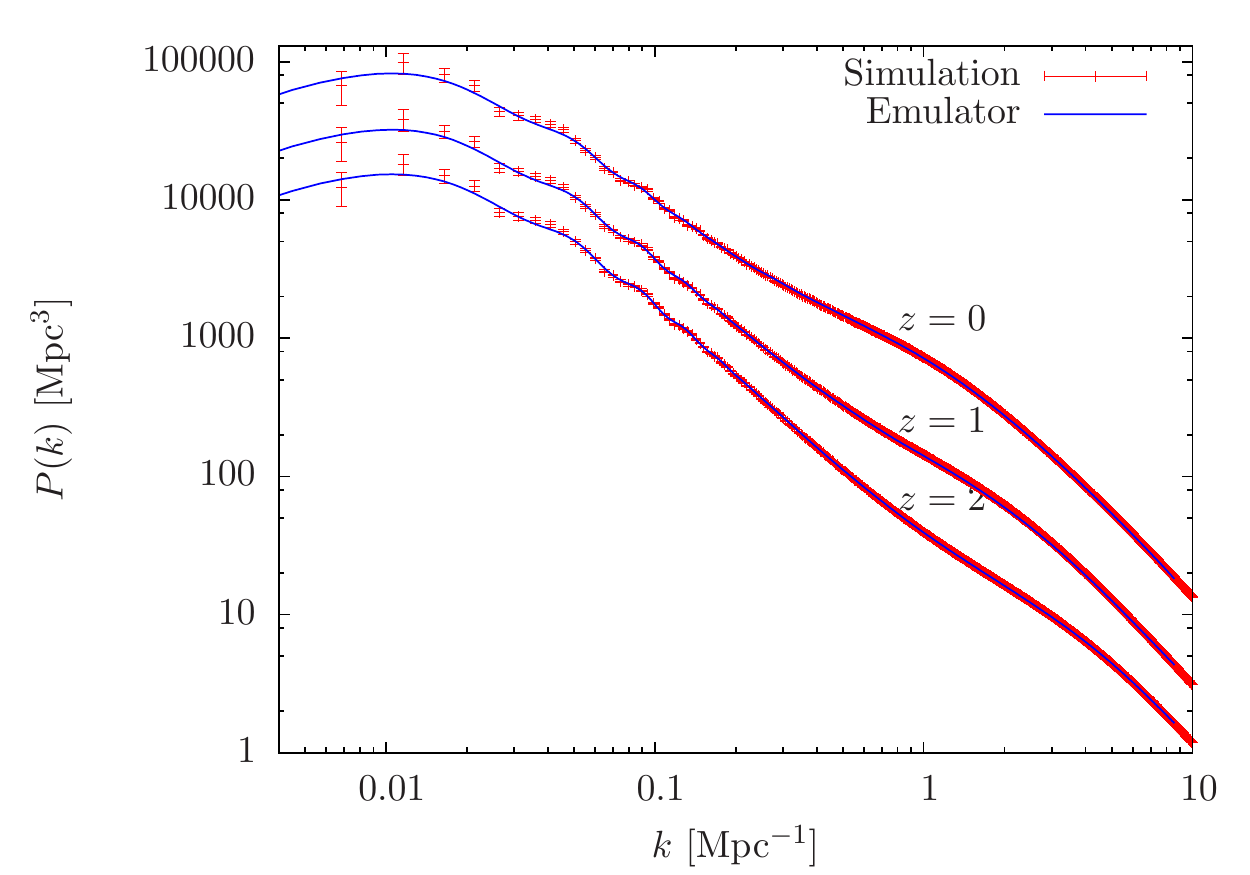}}
\caption{\label{pk}Matter power spectrum measurements at three
  different redshifts, $z=0,1,2$ (top to bottom). Red: Simulation
  outputs, blue: prediction from the extended Cosmic
  Emulator of~\cite{emu_ext}, which allows the Hubble parameter to be
  freely chosen. The corresponding ratios are shown in
  Figure~\ref{pk_rats}. }
\end{figure}

\begin{figure}
\centerline{
 \includegraphics[width=3.7in]{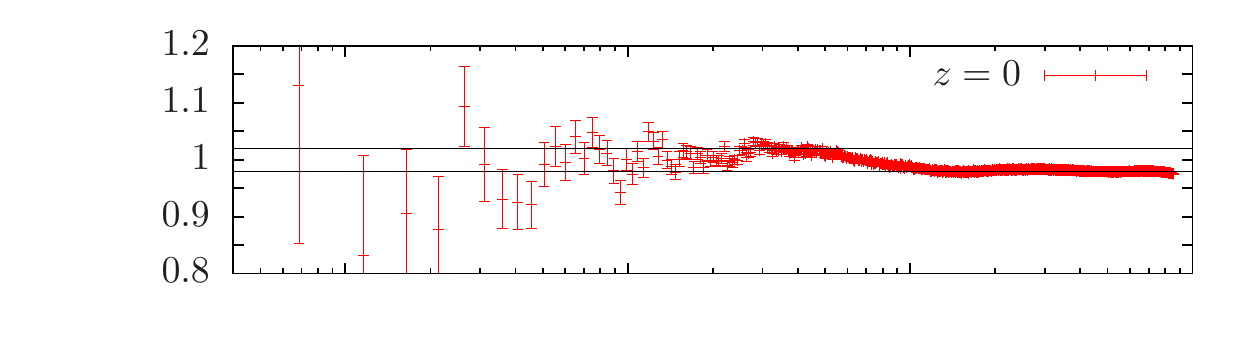}}

\vspace{-0.8cm}

 \centerline{
 \includegraphics[width=3.7in]{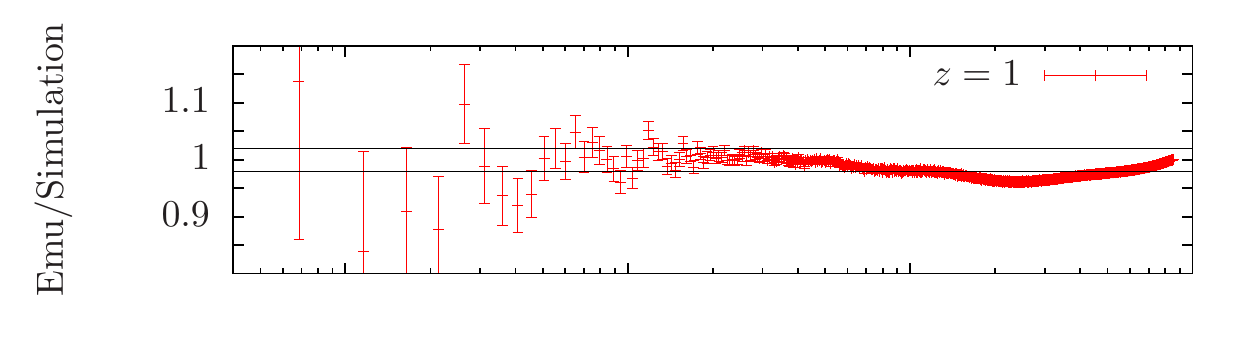}}
 
\vspace{-0.8cm}

 \centerline{
 \includegraphics[width=3.7in]{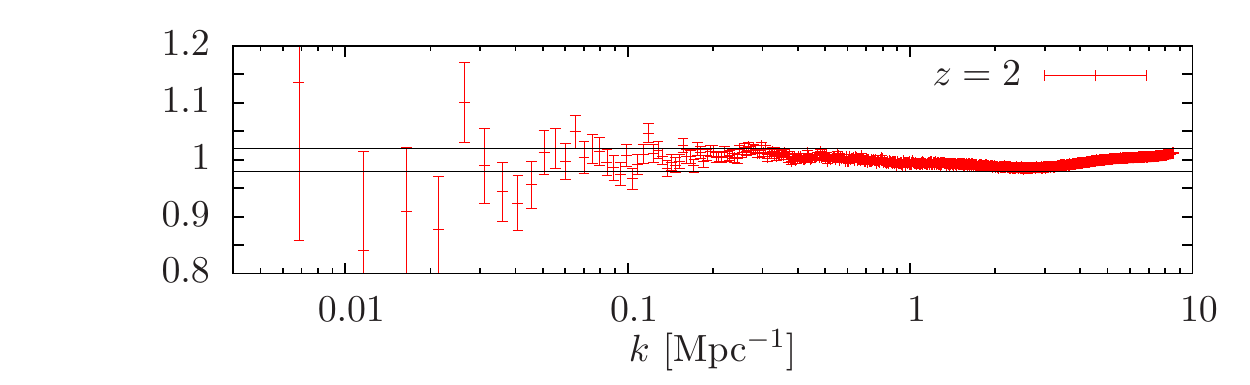}}
 
\caption{\label{pk_rats}Ratios of power spectra measured from the
  simulation with respect to the emulator at three redshifts $z=0,1,
  2$ (top to bottom). The black solid lines in each panel are the
  $\pm2\%$ deviation limits. For redshifts $z=0$ and $z=2$ the
  difference is within 2\% while for $z=1$ it is closer to $4\%$,
  still well within the errors quoted in \cite{emu_ext}.}
\end{figure}

The results for the matter power spectrum at three redshifts are shown
in Figure~\ref{pk}. The error bars are determined from the number of
independent modes within the box. In addition to the power spectra
measurements from the simulation, we also show predictions from the
extended Coyote emulator of \cite{emu_ext}. This emulator was built by
matching simulations of varying box sizes to cover the full $k$-range
of interest -- the high mass and force resolution in a cosmological
volume of the Q Continuum simulation allows us to cover this full
$k$-range with only {\em one} simulation. Figure~\ref{pk_rats} shows
the ratio of the power spectrum at three redshifts with the extended
emulator. The black solid lines in each panel indicate the $\pm2\%$
deviation from the emulator. As originally advertised in
\cite{emu_ext}, the emulator predictions agree with simulation results
well within the 5\% error limit. We note that the agreement with the
larger volume, trillion-particle Outer Rim simulation (run with the
same cosmology, but with a different HACC short-range solver) is at
the level of a fraction of a percent~\citep{habib14}.

\subsection{Halo Mass Function}

We next show results for the halo mass function at three different
redshifts, $z=0, 1$, and $z=2$ as found with an FOF halo finder with a
linking length set to $b=0.2$. Aside from tests of universality, high
accuracy results with this halo definition are available from past
simulation campaigns, covering large simulation volumes by combining
many realizations. Here we show a comparison with the
\cite{bhattacharya11} fit -- for a detailed discussion of how this fit
compares to other contemporary fits the reader is referred to the
original paper.

\begin{figure}
\centerline{
 \includegraphics[width=3.7in]{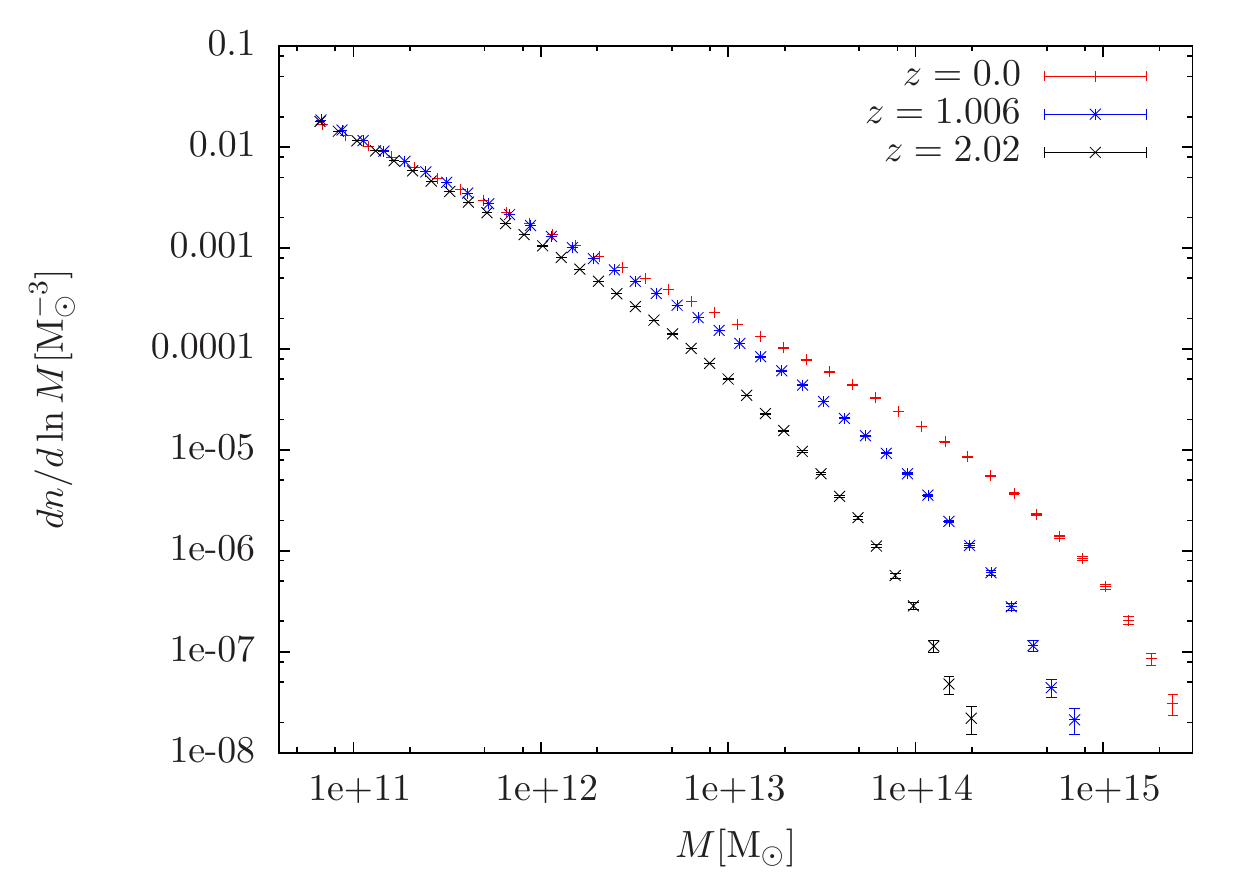}}
\caption{\label{massf}Halo mass function at three redshifts, $z=0,1,2$
  found with an FOF halo finder and linking length $b=0.2$ with at
  least 400 particles per halo. We omit bins with less than 10 halos; in
  the case of $z=2.02$ we therefore drop the last three bins and in
  the case of $z=1.006$ the last two bins.}
\end{figure}

We consider halos that have 400 or more particles, though we have
stored halos with masses down to 100 particles per halo.  The
conservative 400-particle cut was used in \cite{bhattacharya11} to
ensure that the halo mass measurements in the lower mass range are not
biased due to particle under-sampling. In addition to considering only
halos that are well sampled, we also apply a mass correction that
depends on the number of particles in a halo, $n_h$, first suggested
by~\cite{warren} and slightly modified by \cite{bhattacharya11} to
read:
\begin{equation}
n_h^{\rm corr}=n_h(1-n_h^{-0.65}).
\end{equation}     
(The original correction factor used $-0.6$ in the exponent, slightly
overcorrecting the halo masses.) At our halo mass cut of 400 particles,
this leads to a correction of at most $2\%$ for the smallest
halos. \cite{bhattacharya11} also suggested a correction for
insufficient force resolution which in our case is unnecessary due to
our high force resolution. The final result for the mass functions is
shown in Figure~\ref{massf}. Our high mass resolution and relatively
large volume allows us to cover the mass function over a wide range of
halo masses, characteristic of individual galaxies to massive
clusters.

\begin{figure}[t]
\centerline{
 \includegraphics[width=3.7in]{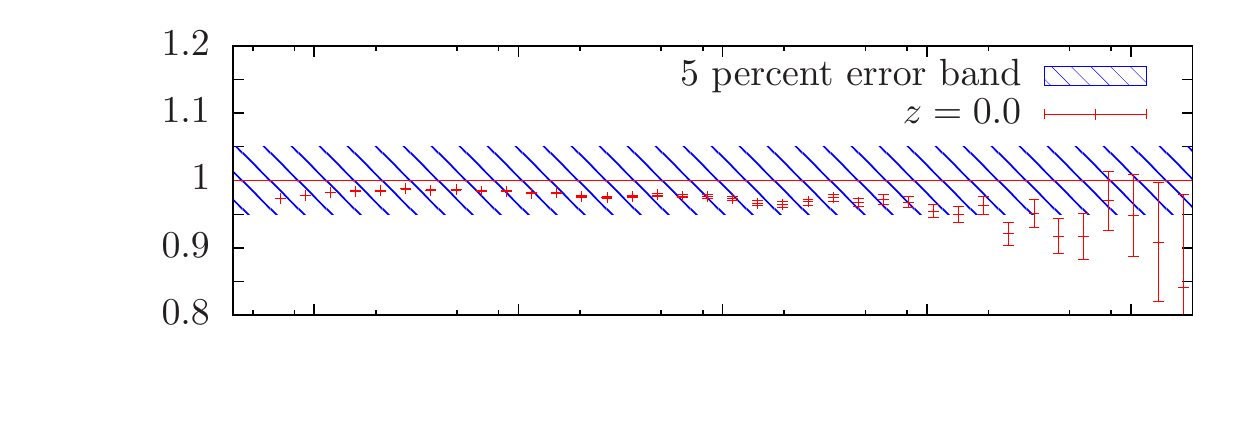}}

\vspace{-1.1cm}

 \centerline{
 \includegraphics[width=3.7in]{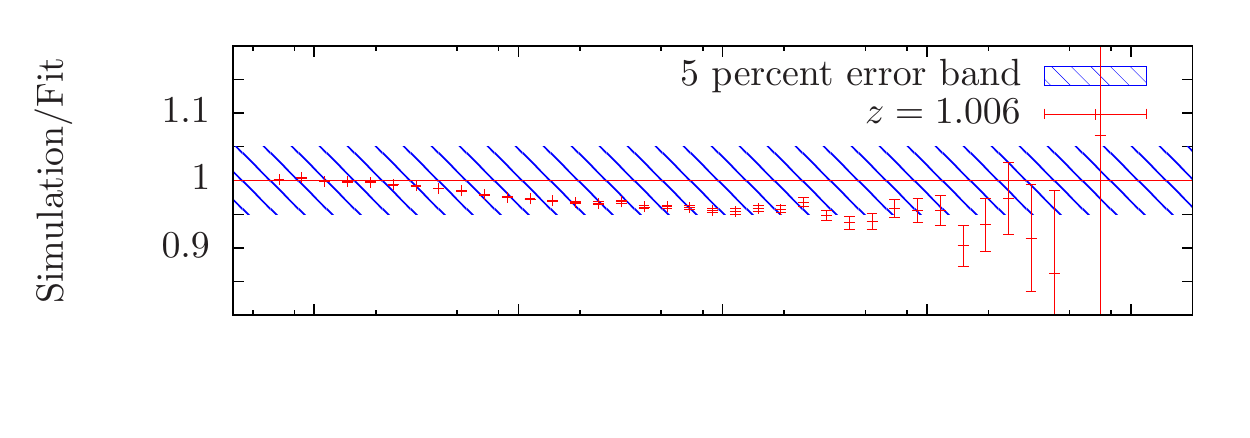}}
 
\vspace{-1.1cm}

 \centerline{
 \includegraphics[width=3.7in]{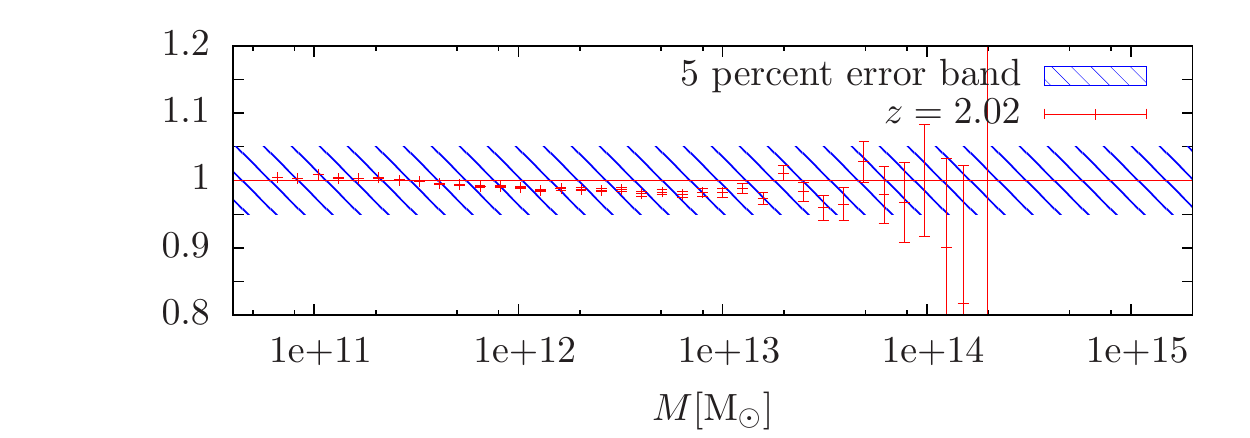}}
 
\caption{\label{massf_rats}Ratios of mass functions measured from the
  simulation with respect to the \cite{bhattacharya11} fit -- derived
  for a slightly different cosmology than that investigated here -- at
  three redshifts $z=0,1, 2$ (top to bottom). The blue hashed region
  in each panel shows the $\pm5\%$ deviation limits. For redshifts
  $z=0$ and $z=1$ the difference is within those limits. For $z=2$ the
  result is very close to the fit.}
\end{figure}

In Figure~\ref{massf_rats} we compare our mass function results with
the fit derived in \cite{bhattacharya11} for the three different
redshifts of interest considered here. The fit is inspired by the
original Sheth-Tormen form~\citep{sheth99}, but has one additional
parameter and allows for an explicit redshift dependence:
\begin{equation}\label{fit1}
f^{\rm Bhatt}(\sigma,z)=A\sqrt{\frac{2}{\pi}}
\exp\left[-\frac{a\delta_c^2}
  {2\sigma^2}\right]\left[1+\left(\frac{\sigma^2}{a\delta_c^2}
  \right)^p\right] \left(\frac{\delta_c\sqrt{a}}{\sigma}\right)^q,
\end{equation} 
with the following parameters (note that all parameters are different
from the original Sheth-Tormen choices): 
\begin{equation}\label{fit2}
A=\frac{0.333}{(1+z)^{0.11}};~~a=\frac{0.788}{(1+z)^{0.01}};~~p=0.807,~~q=1.795.  
\end{equation}
As was pointed out in \cite{bhattacharya11} the change of the density
threshold for spherical collapse, $\delta_c$, with redshift does not
improve the quality of the mass function fit. We therefore also keep
$\delta_c=1.686$ for all redshifts.

The necessity of the redshift dependent terms in the mass function fit
underline results from previous work: universality of the mass
function is broken at the $\sim 5-10\%$ level even within the same
cosmology if different epochs are considered.

\begin{figure}[t]


 \centerline{
 \includegraphics[width=3.7in]{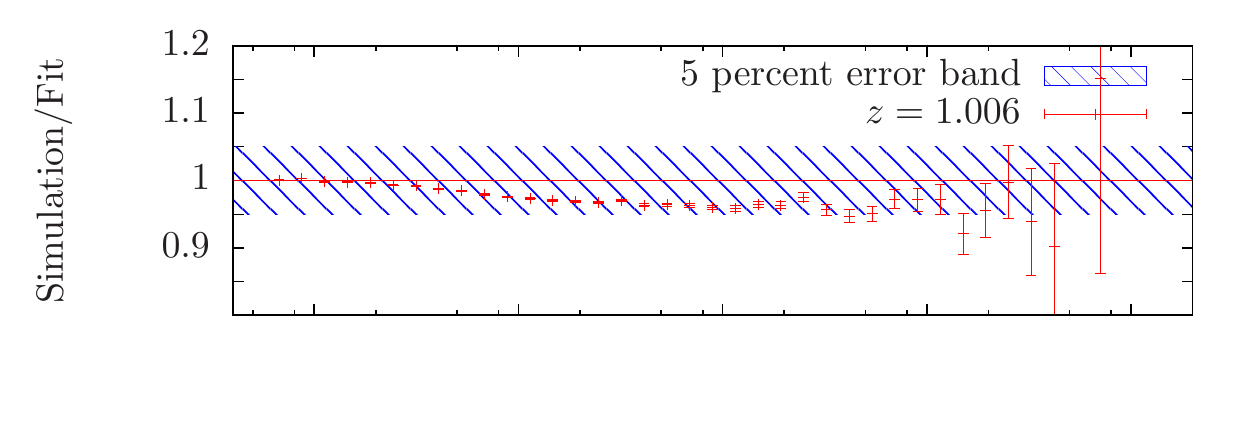}}
 
\vspace{-1.1cm}

 \centerline{
 \includegraphics[width=3.7in]{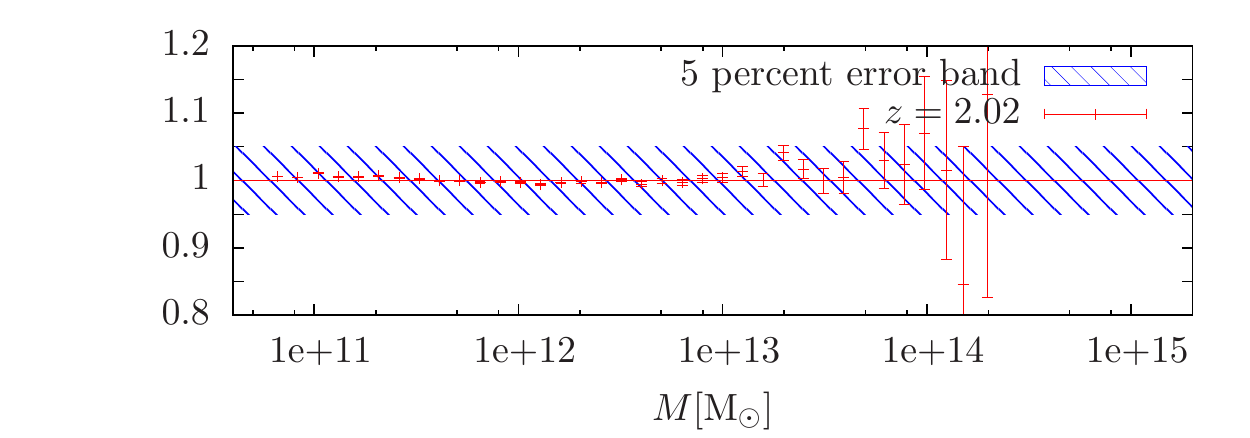}}
 
\caption{\label{massf_rats_noz}Same as in Figure~\ref{massf_rats} for
  $z=1$ and $z=2$ but only keeping the redshift dependence in the
  amplitude $A$ for the mass function fit of Eq.~(\ref{fit2}),
  i.e. simply using $a=0.788$. The result for $z=0$ is unchanged. For
  $z=2$ the agreement between the fit and the simulation is
  essentially perfect; for $z=1$ we see a slight improvement over the
  original fit.}
\end{figure}

In comparison to~\cite{bhattacharya11}, the total volume is
considerably smaller (\citealt{bhattacharya11} had approximately 100
times the volume we cover here), but the mass resolution is
significantly better (by a factor of 10, even in the worst case). The
improved mass resolution combined with a large volume allows us to
cover the mass function at higher redshifts over a wide mass range --
as can be seen in Figure~\ref{massf}; even at $z=2$ the mass range
extends from $10^{11}$~M$_\odot$ to $2\cdot 10^{14}$~M$_\odot$. This
increase in dynamic range over previous work allows us to
reinvestigate the $z$-dependence of the mass function as introduced
in~\cite{bhattacharya11}. We find that an explicit $z$-dependence
introduced into the shape of the mass function via the fitting
parameter $a$ is in fact not needed (at least not for the redshift
range considered here). We show the ratio of our mass function with
respect to a fit that simply uses $a=0.788$ in
Eqs.~(\ref{fit1},\ref{fit2}) improves the agreement between fit and
simulation. The good level of agreement between our results and those
presented in~\cite{bhattacharya11} showcases the overall consistency
of results obtained from three independent cosmological N-body codes
({\sc Gadget-2}, HACC, and TreePM).
 
\subsection{Concentration Mass Relation}

Finally, we show results for the halo concentration-mass relation,
measured at two different redshifts. This relation has been recently
measured for clusters by the CLASH collaboration~\citep{clash}, the
LoCuSS project~\citep{okabe}, and by~\cite{newman}. All of these
results are in very good agreement with the theoretical predictions of
\cite{bhattacharya13}. At smaller mass scales, the concentration-mass
relation can be determined by galaxy-galaxy lensing
studies~\citep{mandelbaum}.

We measure $M_{200}$ for all halos with
at least 1000 particles in the simulation and determine for each halo
its best-fit Navarro-Frenk-White (NFW) profile
\citep{navarro96,navarro97}. In the following, both $M_{200}$ and the
corresponding concentration $c_{200}$ denote measurements with respect
to critical overdensity. Here we do not classify the halos into
relaxed and unrelaxed, but consider the entire sample.  We follow the
same procedure as outlined in \cite{bhattacharya13} to determine the
$c-M$ relation.

\begin{table}
\caption{Power law fit parameters at $z=0$ from different
  simulations\label{tab1}} 
\begin{tabular}{cccccc}
$\alpha$ & $\beta$ & $\Omega_m$ & $\sigma_8$ & $h$ & Reference\\ 
\hline\hline
6.34 & -0.097 & 0.258 & 0.796 & 0.719 & Duffy 08 \\
6.35 & -0.11 & 0.258 & 0.796 & 0.72 & Macci{\`o} 08\\
7.02 & -0.08 & 0.25 & 0.8 & 0.72 & Bhattacharya 13\\
7.20 & -0.0926 & 0.265 & 0.8 & 0.71 & this paper \\
\end{tabular}
\end{table}

In the past, many studies have been carried out to determine the shape
of the concentration-mass relation (see, e.g.,
\citealt{bullock99,eke01,gao07,maccio07,neto07,duffy08,hayashi08,
  zhao09,prada11,klypin,bhattacharya13,dutton14,diemer14}). In most
cases it was found that a simple power law fit works well to describe
the halo concentration as a function of mass: 
\begin{equation}
c_{200}=\alpha\left(\frac{M_{200}}{10^{12}{\rm M}_\odot}\right)^\beta,
\end{equation}
where $M_{200}$ is measured in M$_\odot$. One drawback of this
description is that it does not capture any redshift dependence and
the best-fit power law has to be determined for each redshift
separately. For this reason, several attempts were made to provide a
more general fit that does include the redshift dependence in a closed
form.  As \cite{gao07} point out, in particular the early attempts to
provide such a fit do not reproduce large, high-resolution simulation
results well, and for very accurate $c-M$ predictions, power-law fits
tuned to different redshifts perform much better.  On the other hand,
\cite{bhattacharya13} found that instead of using the $c-M$ relation
directly, a $c-\nu$ relation with $\nu=\delta_c(z)/\sigma(M,z)$
delivers an expression that is approximately constant over a redshift
range $z=0 - 2$ and is therefore well suited to describe the redshift
evolution in a closed form. In a recent paper, \cite{kwan13}
introduced a $c-M$ emulator to provide predictions for the $c-M$
relation for any redshift between $z=0$ and $z=1$ over a range of
$w$CDM cosmologies. The emulator was based on a large set of
simulations from the Coyote Universe suite~\citep{coyote3}.

\begin{figure}[t]
\centerline{
 \includegraphics[width=3.7in]{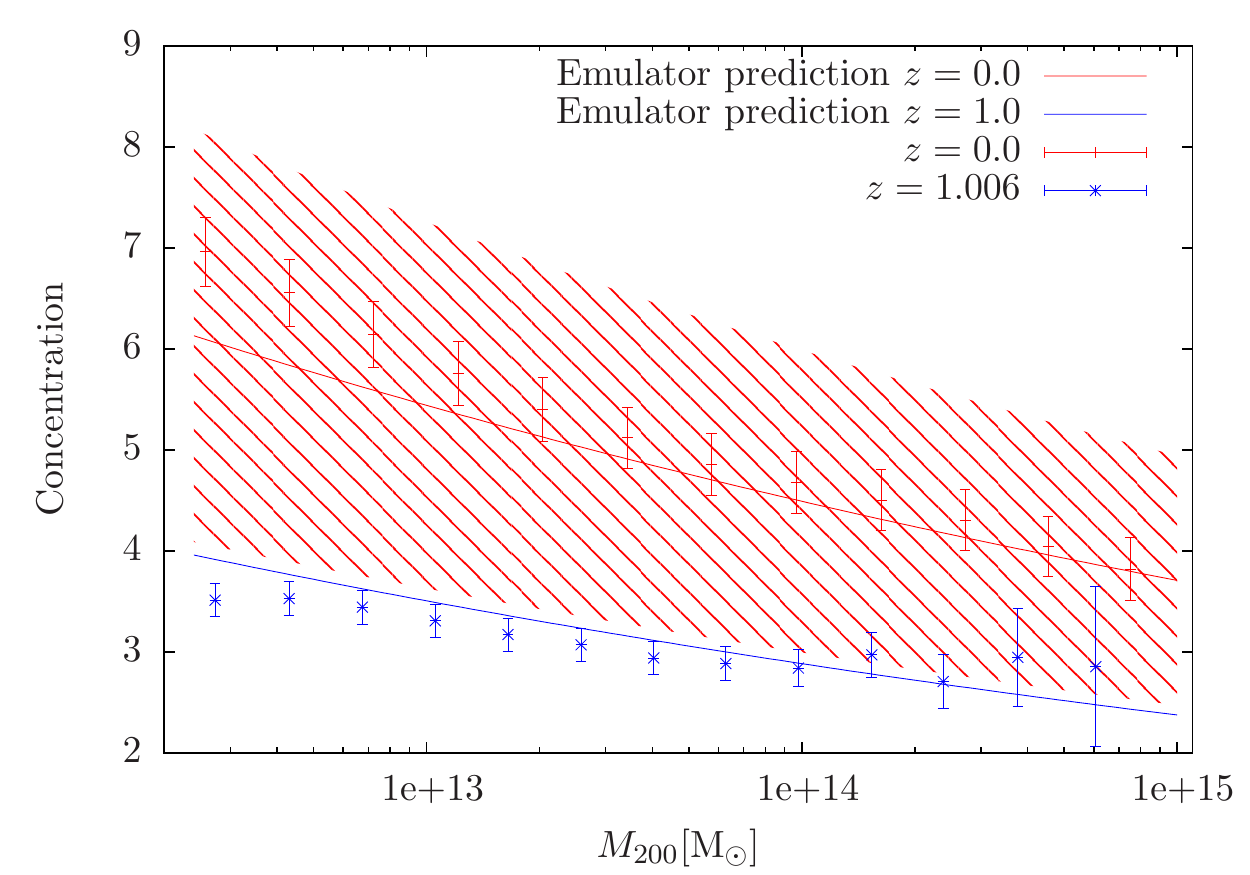}}
\caption{\label{cM}Concentration-mass relation at redshifts,
  $z=1$ and $z=2$ from the simulation (points with error bars) and
  the predictions from the \cite{kwan13} emulator (the emulator does
  not provide results at lower masses or beyond $z=1$). The hashed
  region depicts the intrinsic scatter in the concentration-mass
  relation.} 
\end{figure}

\begin{figure}[t]
\centerline{
 \includegraphics[width=3.7in]{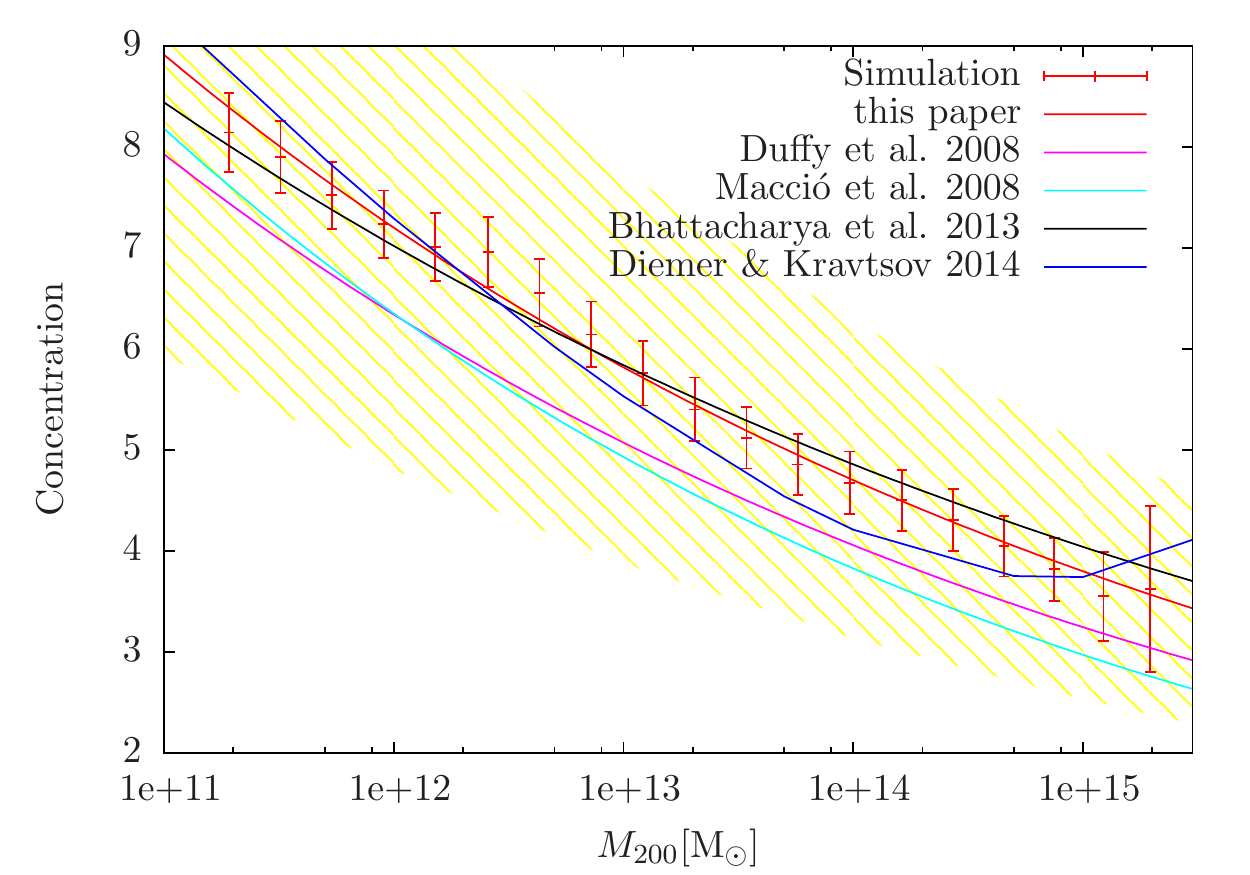}}
\caption{\label{cMfits}Concentration-mass relation over the full mass range covered
 by the Q Continuum simulation at redshift $z=0$ (points with error
  bars) and the predictions from various groups. The yellow shaded region shows the intrinsic
  scatter. All predictions and the simulation results are well within that scatter.}
\end{figure}

Figure~\ref{cM} shows our concentration mass measurements at two
redshifts ($z=0,1$) and the corresponding emulator predictions. Our
measured results agree well with the emulator prediction, which exist
only over a limited mass and redshift range. In Figure~\ref{cMfits} we
compare our results at $z=0$ with several other proposed fits obtained
from various simulations. The hashed region in both figures indicates
the intrinsic scatter in the concentration-mass relation. We emphasize
that this is the first time that the $c-M$ relation has been measured
from a single simulation volume over such an extended mass range; the
large volume eliminates possible concerns about biases due to too
small simulation boxes. In Figure~\ref{cMfits} we restrict our
comparison to results that are close to the cosmology we consider in
this paper. Most of the fits are based on a power law assumption; we
give the exact parameters as well as the cosmologies simulated in
Table~\ref{tab1}. \cite{diemer14} provide a more general prediction
(shown in dark blue) for the $c-M$ relation, displaying a turn-up at
large masses. We do not observe such a behavior, although as discussed
in detail in \cite{dutton14} such an upturn could depend on how the
concentration is determined. The yellow hashed region in
Figure~\ref{cMfits} shows the estimated intrinsic scatter in the
concentration mass relation -- all the predictions fall well within
this band.  The prediction from \cite{maccio08} and \cite{duffy08} are
somewhat lower than what we obtain, in agreement with the previous
findings of \cite{bhattacharya13}.

\section{Conclusions and Outlook}
\label{conclusion}
In this paper we have introduced the Q Continuum simulation, the
largest cosmological N-body simulation at a mass resolution of
$m_p\sim 1.5\cdot 10^8$~M$_\odot$ carried out so far. We have described
the parameters of the simulation and of its outputs, giving a detailed
description of the raw particle data files to halo catalogs and power
spectra. These outputs can be used for a very large range of
cosmological applications.

We have assembled a set of first measurements of the matter power
spectrum, the mass function, and the concentration-mass relation,
showing that the results agree well with previous predictions in the
literature.  This is the first time that these predictions are
obtained down to small halo mass ranges from a single simulation with
box size in the $\sim$Gpc range, thus eliminating a number of concerns
about biases due to small volume effects and to errors when matching
results across multiple box sizes.

The outputs generated by this simulation are currently stored and have
a data volume of more than 2~PB. A fast merger tree algorithm that
includes an inbuilt subhalo finder is a key component of the analysis
suite (Rangel et al., in preparation). Various analyses are currently
in progress, including a suite of galaxy catalogs, based on the
semi-analytic modeling code Galacticus as well as using a S/HAM
approach that works directly with the merger trees.

We are tuning these approaches on smaller test simulations (at similar
mass and force resolution but reduced volume) and will be able to
implement them immediately once the run of the analysis suite has
completed. The simulation will be also very valuable for diverse
lensing measurements. Scalable tools have been set up for measurements
such as strong lensing, weak lensing shear, cluster (strong and weak)
lensing, and galaxy-galaxy lensing. In particular, for strong lensing
and galaxy-galaxy lensing applications, the high-mass resolution and
the ability to resolve small subhalos is very important. 

\begin{figure}[t]
\centerline{
 \includegraphics[width=3.5in]{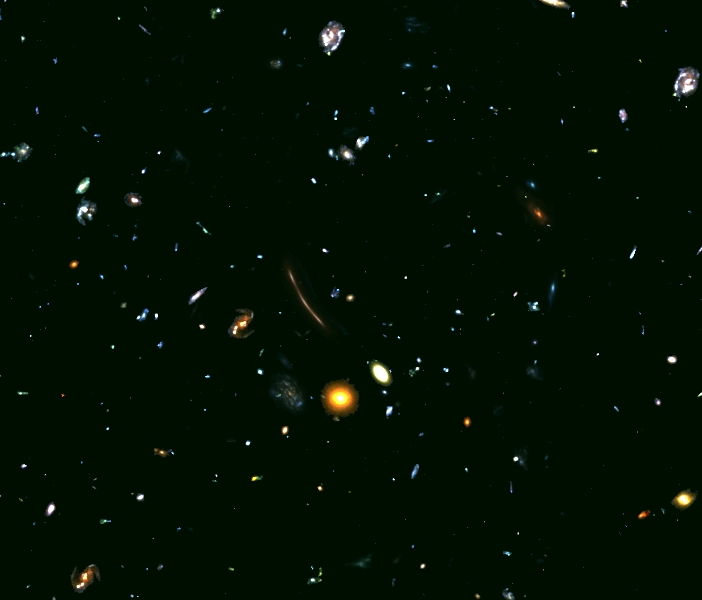}}
\caption{\label{lens}Strong lensing arc generated from the
  cluster-scale halo shown in Figure~\ref{albert}.  Source and
  foreground galaxies are placed using galaxy images from the Hubble
  Ultra Deep Field. This image does not include noise, PSF
  convolution, and cluster galaxies for clarity, though these
  capabilities are opart of the lensing pipeline.}
\end{figure}

Figure~\ref{lens} shows an example from our strong lensing pipeline --
the arc in the center of the image is due to lensing by the halo shown
in Figure~\ref{albert}. The background and source galaxy images are
taken from the Hubble Ultra Deep Field. The full lensing pipeline
includes the effects of noise, PSF convolution, and methods for
incorporating cluster galaxies.  By combining very high resolution
simulations with the best data available it is possible to generate
synthetic skies very close to reality.

With its combination of large volume and high mass resolution, the Q
Continuum run will be a valuable resource for precision cosmological
studies of large-scale structure formation as well as a testbed --
providing both theoretical predictions and synthetic sky catalogs for
end-to-end survey analyses -- for surveys such as DES, DESI, and
LSST. While we lack the resources to make the full raw particle
outputs available to the community, we have an extensive plan to make
the analyzed data products and resulting synthetic sky catalogs
publicly accessible in the near future.

\begin{acknowledgments}

  Argonne National Laboratory's work was supported under the
  U.S. Department of Energy contract DE-AC02-06CH11357. Partial
  support for HACC development was provided by the Scientific
  Discovery through Advanced Computing (SciDAC) program funded by the
  U.S. Department of Energy, Office of Science, jointly by Advanced
  Scientific Computing Research and High Energy Physics. N. Frontiere
  acknowledges support from the DOE CSGF Fellowship program. This
  research used resources of the Oak Ridge Leadership Computing
  Facility (OLCF) at the Oak Ridge National Laboratory, which is
  supported by the Office of Science of the U.S. Department of Energy
  under Contract No. DE-AC05-00OR22725. The work presented here
  results from an award of computer time provided by the Innovative
  and Novel Computational Impact on Theory and Experiment (INCITE)
  program at the OLCF. We thank Mike Gladders, Michael Florian, Nan Li, 
  and Steve Rangel for the strong lensing image shown in the paper.
  We are indebted to Jack Wells and the OLCF team
  for their outstanding support in enabling the simulation.

\end{acknowledgments}

\end{document}